\documentclass[11pt]{article}
\pdfoutput=1

\usepackage{aas_macros,amsmath,amssymb,comment,cite,esint,graphicx,mathtools}
\usepackage[margin=.8in,letterpaper]{geometry}
\usepackage[colorlinks=false]{hyperref}
\usepackage[affil-it]{authblk}
\usepackage{url}
\usepackage{color}

\numberwithin{equation}{section}
\newcommand{\be}{\begin{equation}}
\newcommand{\ee}{\end{equation}}
\newcommand{\bea}{\setlength\arraycolsep{2pt} \begin{eqnarray}}
\newcommand{\eea}{\end{eqnarray}}
\newcommand{\nn}{\nonumber}

\newcommand{\orr}{\omega_{\mathrm{orb}}}
\newcommand{\pr}{\omega_{\mathrm{prec}}}
\newcommand{\fp}{\mathrm{UFPO}}
\newcommand{\Q}{\mathcal{Q}}
\newcommand{\E}{\mathcal{E}}
\newcommand{\LL}{\mathcal{L}}
\newcommand{\R}{\mathcal{R}}

\makeatletter

\newcommand{\Rmnum}[1]{\expandafter\@slowromancap\romannumeral #1@}
\makeatother

\newcommand\tbbint{{-\mkern -16mu\int}}

\newcommand\dbbint{{-\mkern -19mu\int}}

\newcommand\bbint{
{\mathchoice{\dbbint}{\tbbint}{\tbbint}{\tbbint}}
}

\def\mc{\mathcal}

\def \nn {\nonumber}

\begin{document}
\title{\textbf{Correspondence of eikonal quasinormal modes and unstable fundamental photon orbits for Kerr-Newman black hole}}

\author{
Peng-Cheng Li$^{1,2}$, Tsai-Chen Lee$^{2}$, Minyong Guo$^{1\ast}$,
Bin Chen$^{1,2,3}$}
\date{}

\maketitle

\vspace{-10mm}

\begin{center}
{\it
$^1$Center for High Energy Physics, Peking University,
No.5 Yiheyuan Rd, Beijing 100871, P. R. China\\\vspace{4mm}

$^2$Department of Physics, Peking University, No.5 Yiheyuan Rd, Beijing
100871, P.R. China\\\vspace{4mm}

$^3$ Collaborative Innovation Center of Quantum Matter,
No.5 Yiheyuan Rd, Beijing 100871, P. R. China\\\vspace{2mm}
}
\end{center}

\vspace{8mm}

\begin{abstract}
 In this work, we study the relation of the eikonal quasinormal modes (EQNMs)  and the unstable fundamental photon orbits (UFPOs) in the Kerr-Newman spacetime. We find that in the eikonal limit the gravitational and electromagnetic perturbations of the Kerr-Newman black hole are naturally decoupled, and a single one-dimensional Schr\"odinger-like equation encoding the QNM spectrum can be derived. We then show that the decoupled Teukolsky master equation and the Klein-Gordon equation for the massless scalar field in the Kerr-Newman spacetime are of the same form in the eikonal limit. As a direct consequence, taking into account of the boundary conditions for EQNMs we show an exact correspondence between EQNMs and UFPOs, that is, EQNM/UFPO correspondence. More precisely, similar to the Kerr case, the real part of EQNM's frequency is a linear combination of the  precessional and (polar) orbital frequencies, while the imaginary part of the frequency is proportional to the Lyapunov exponent of the UFPO.

\end{abstract}

\vfill{\footnotesize Email: lipch2019@pku.edu.cn,\,tsaichen\_lee@pku.edu.cn,\,minyongguo@pku.edu.cn,\,bchen01@pku.edu.cn.\\$~~~~~~\ast$ Corresponding author.}

\maketitle

\newpage
\baselineskip 18pt
\section{Introduction}\label{sec:intro}
According to the unique theorems \cite{Chrusciel:2012jk}, in four-dimensional spacetime, the most general stationary asympotically flat black hole  solution to the electro-vacuum Einstein field equations is the Kerr-Newman (KN) black hole \cite{Newman:1965my,Adamo:2014baa}. The solution is uniquely  characterized by the mass $M$, angular momentum $J=M a$ and the charge $Q$. The Kerr, Reissner-Nordstrom (RN) and Schwarzschild black holes correspond to the limiting cases of the KN black hole: $Q=0$, $a=0$ and $a=Q=0$, respectively. Although it is believed that the astrophysical black holes are electrically neutral \cite{Gibbons:1974zd,Bozzola:2020mjx,Wang:2020fra,Wang:2021uuh}, charged black holes are still of great interest in several aspects. For example, the stability of the perturbed KN black hole is still a major unsolved problem in General Relativity \cite{Dias:2015wqa}.


When a KN black holes is perturbed, the linear perturbations are composed of a set of  characteristic modes that satisfy an ingoing boundary condition at the horizon and an outgoing boundary condition at infinity. These oscillatory and decaying modes are called the quasinormal modes (QNMs) \cite{Kokkotas:1999bd,Berti:2009kk,Konoplya:2011qq}, which play an important role in the study of black holes. For example, the complex frequencies of the QNMs can be used to determine the linear stability of a perturbed BH. Moreover, in the ringdown stage of the coalescence of two astrophysical
black holes, the gravitational waves (GWs) take the form of superposed QNMs of the remnant black hole. As a consequence of the no-hair theorem \cite{Bekenstein:1996pn}, the measurement of the QNMs would help us to test general relativity and probe the  nature of remnants from compact binary mergers \cite{Berti:2018vdi}. In general, the calculation of QNM relies on the separation of the linear perturbations in all variables. The QNM spectrum appears  then as the eigenvalues of a single one-dimensional Schr\"odinger-like equation. This procedure is achievable for the Schwarzschild, the RN and the Kerr black holes. For the Kerr black hole, such an equation is known as the Teukolsky equation \cite{Teukolsky:1973ha}. However, it does not seem possible to cast the general perturbations of a KN black hole into a single equation, due to the coupling between different kind of perturbations. Except for some limiting cases, such as the weakly charged \cite{Mark:2014aja} or slowly rotating cases \cite{Pani:2013ija,Pani:2013wsa},  one has to resort to  numerical technique to handle the coupled partial differential equations in order to calculate the QNMs \cite{Berti:2005eb,Dias:2015wqa}.

On the other hand, null geodesics have been studied extensively in various black hole backgrounds, and many special optical characteristics were found in the presence of a black hole. As pointed out in \cite{Bardeen:1973tla,Chandrasekhar:1985kt}, the gravity around a black hole is strong enough that the light would bend very strongly so that the photons under certain conditions would move along the bounded spherical orbits, which are called fundamental photon orbits (FPOs)\footnote{In the literatures, spherical photon orbits (SPOs) are sometimes used to denote the photon orbits with a constant radius in the Boyer-Lindquist coordinates of Kerr and KN spacetimes instead of FPOs. However, it is imprecise to use the term ``SPOs", since $r=\mbox{const.}$ does not really correspond to a sphere in Boyer-Lindquist coordinates. To avoid the ambiguity, we would like to use FPOs rather than SPOs in our paper. A rigorous definition of FPOs are be found in \cite{Cunha:2017eoe}}\cite{Cunha:2017qtt}.  For a general stationary axisymmetric black hole spacetime, such orbits could be stable or unstable in the radial direction of FPOs \cite{Cunha:2017eoe, Cunha:2020azh,Guo:2020qwk}. However, the unstable FPOs (UFPOs) are of more concern since under a slight perturbation, they would either fall into the black hole or escape to the infinity. The photons in the  ``nearly bounded'' UFPOs could enter the eyes of distant observers away from the black holes. The radii and impact parameters of these UFPOs are found to be confined to a certain range. The radial deviations from the UFPOs turn out to be exponentially increasing, and the exponential factor is referred  to as the Lyapunov exponent \cite{Ferrari:1984zz,Cardoso:2008bp}. Moreover, there has been some works that tried to  build connections between FPOs and thermodynamics of the black holes, see \cite{Zhang:2019glo, Li:2019dai, Xu:2019yub, Zhang:2019tzi, Han:2018ooi, Wei:2018aqm, Cai:2021fpr}.

In particular, QNMs and geodesic photon orbits (GPOs),  the two seemingly very different things are actually closely related. It was found that for the Schwarzschild, the RN and the Kerr spacetimes, the eikonal QNMs (EQNMs) of the gravitational perturbations correspond to the specific null geodesics that reside on the spherical photon orbits, or  the UFPOs \cite{Ferrari:1984zz,Cardoso:2008bp,Yang:2012he}. Initially, Ferrari and Mashhoon \cite{Ferrari:1984zz} showed the QNM frequency of perturbed Schwarzschild black holes in the eikonal limit has a very close connection with the Keplerian frequency of the circular photon orbit and Lyapunov exponent of the orbit. In addition, they found similar results for slowly rotating black holes. In the sequent works, Cardoso et al. \cite{Cardoso:2008bp} generalized the correspondence to the stationary, spherically symmetric and asymptotically flat spacetimes in any dimensions. Later on, by comparing the  WKB calculation of the Teukolsky equation in the eikonal limit and the Hamilton-Jacobi equations in the Kerr spacetime,  Yang et al.  \cite{Yang:2012he} found a relationship between the EQNM frequencies of Kerr black holes of arbitrary spins and UFPOs. More precisely they showed that when $l\gg1$, the QNM frequencies $\omega=\omega_R-i\omega_I$ can be written as
\be\label{QNMGeo}
\omega=\left(l+\frac12\right)\left(\orr+\frac{m}{l+\frac12}\pr\right)-i\left(n+\frac{1}{2}\right)\gamma_L,
\ee
where $\orr$ is the frequency at which the photon oscillates below and above the equatorial plane, $\pr$ is the Lense-Thirring precession
frequency and $\gamma_L$ is the Lyapunov exponent of the spherical photon orbit. Moreover, $l$, $m$ are the familiar angular multipoles and $n$ is the overtone number. Due to the relation of FPOs and black hole shadow \cite{Bardeen:1973tla,Chandrasekhar:1985kt}, the QNM/geodesic correspondence can be used to relate EQNMs with the black hole shadows, see the recent works \cite{Stefanov:2010xz,Jusufi:2019ltj,Cuadros-Melgar:2020kqn,Yang:2021zqy}.

In this paper, we would like to investigate whether the EQNM/UFPO correspondence reviewed above (\ref{QNMGeo}) is valid for the KN black holes\footnote{Note that unless specified, we always refer to the QNMs of gravitational perturbations.}. Before we proceed to this goal, we would  point out a simple but essential fact that would be useful as we explore the EQNM/UFPO correspondence for the KN black holes. In the eikonal limit, or equivalently the high frequency limit, both the electromagnetic and massless scalar waves  behave like massless particles moving along null geodesics in the general curved spacetime. Since the scalar QNMs can be viewed as the waves propagating in the black holes background with proper boundary conditions, it is expected that in the eikonal limit, the scalar QNMs correspond to some special null geodesics, whose form depends on the boundary conditions being considered. Therefore, the EQNM/UFPO correspondence (\ref{QNMGeo}) is possible only when the QNM equation can be transformed into the massless Klein-Gordon equation, otherwise the elegant relation would be broken. For example, for asymptotically flat black holes in the Einstein–Lovelock gravity, it was found \cite{Konoplya:2017wot} that  all three types of perturbations satisfy the equations different from the (separated) massless Klein-Gordon equation,  indicating the violation of the correspondence (\ref{QNMGeo}). 


Now let us get back to the KN black hole. Shortly after the pioneer work \cite{Ferrari:1984zz}, Mashhoon studied the linear stability of KN black holes via the QNMs obtained from the EQNM/UFPO correspondence, which has not yet been proven to be valid for the KN black holes \cite{Mashhoon:1985cya}. By following the work \cite{Yang:2012he}, Zhao et al. \cite{Zhao:2015pqa} built a relation between the QNMs of a test charged scalar field and the (modified) geodesics in the KN spacetime\footnote{Due to the presence of Lorentz force, the charged particles no longer move along the geodesics.}. However, it is known that the QNMs of a scalar field is significantly different from those from gravitational perturbations. Thus, obtaining the analog of the Teukolsky equation  for the KN black holes is a prerequisite for exploring the EQNM/UFPO correspondence. Due to the inseparability of the coupling between the gravitational perturbations  and  electromagnetic  perturbations, to date all attempts to cast the general perturbations of a KN black hole into a single differential equation have failed \cite{Chandrasekhar:1985kt}. However, as we will show in this work, the gravitational perturbation are naturally decoupled from the electromagnetic perturbations in the eikonal limit, such that the analog of the Teukolsky equation for the KN black hole can be derived as well. We further show that similar to the Kerr case, this equation is equivalent to the (separated) massless Klein-Gordon equation. This equivalence suggests that EQNMs must have a definite correspondence with GPOs. Next, considering the boundary conditions of EQNMs, we can further identify that the GPOs corresponding to EQNMs is nothing but UFPOs, and we establish the EQNM/UFPO correspondence for the KN black holes.

The remaining parts of the paper is organized as follows.  In section \ref{QNMMSWN}, we present a detailed study of the perturbations of the KN black holes in the eikonal limit. In section \ref{QNMG}, we prove the EQNM/UFPO correspondence for KN black holes by considering the boundary conditions of EQNMs. We summarize and discuss our results in section \ref{CD}. In section \ref{EWMSW}, we give a quick review of the geometric optics approximation in a curved spacetime, and show the equivalence between the Teukolsky equation and the (separated) massless Klein-Gordon equation in the eikonal limit for the Kerr black holes.

\section{ Eikonal QNM for KN black holes}\label{QNMMSWN}

In this section we study the perturbation equations of the charged black holes in the eikonal limit. For the static Reissner-Nordstrom black holes, since the black holes are charged, purely electromagnetic perturbations induce gravitational perturbations and vice versa. In  this case the gravitational and electromagnetic perturbations are coupled together, which makes the separation in $r$ and $\theta$ difficult. Fortunately, due to the symmetry of the spacetime, Moncrief and Zerilli successfully decoupled the perturbation equations by considering  the linear combination of the gravitational and electromagnetic perturbations, i.e. the so-called ``gravito-electromagnetic'' perturbations \cite{Moncrief:1974ng,Moncrief:1974gw,Zerilli:1974ai}. With this the complete separation becomes possible and the QNMs can be calculated \cite{Berti:2009kk}.

For the Kerr-Newman black holes, in contrast to Kerr black holes or RN black holes, to date all attempts to decouple the electromagnetic and gravitational perturbations have failed. For example, Dudley and Finley obtained approximate decoupled equation (dubbed DF equation) describing the propagation of spin-weighted test fields the Kerr-Newman spacetime \cite{Dudley:1977zz,Dudley:1978vd}. However, the DF equation was derived under the assumption that the electromagnetic and gravitational perturbations of KN black holes could be treated independently, which is only a rough approximation \cite{Berti:2005eb}. Furthermore, as shown by \cite{Mark:2014aja} the DF equation can be understood as the Teukolsky equation with modification $\Delta_{Kerr}\to\Delta_{KN}$. It is clearly then the DF equation correctly captures the QNMs of a massless scalar field in the KN spacetime. But for other kinds of perturbation, they are coupled with each other and cannot be separated in general. For the recent developments of the QNMs of the KN black holes, one can see
\cite{Pani:2013ija,Pani:2013wsa,Mark:2014aja,Dias:2015wqa,Zilhao:2014wqa,Hod:2014uqa,Hod:2015xlh,Zimmerman:2015trm,Wang:2021uuh}. However, as we will see below, in the eikonal limit, significant simplification occurs and the gravitational perturbations decouple naturally from the electromagnetic ones, and moreover  the complete separation in all variables becomes viable.

In terms of the Boyer-Lindquist coordinates, the metric of the KN spacetime is of the form
\be
  ds^2 = -d t^2 + {{\Sigma}\over \Delta_{KN}}d r^2 +\Sigma d\theta^2
  +(r^2+a^2)\sin^2\theta\,d\phi^2 +{{2Mr}\over{\Sigma}}
  (a\sin^2\theta\,d\phi - d t)^2\ ,\label{7.114}
\ee where \be
  \Delta_{KN}(r) = r^2 - 2Mr +a^2+Q^2\ ,
\hspace{3ex}  \Sigma(r,\theta) = r^2+a^2\cos^2\theta\ .\label{7.116}
\ee
When $Q=0$, it reduces to the metric of the Kerr spacetime.
As the Kerr spacetime, the KN spacetime is also of type D in the Petrov classification, which indicates that the Weyl scalars, $\Psi_0$, $\Psi_1$, $\Psi_3$ and $\Psi_4$, and the spin coefficients, $\kappa$, $\sigma$, $\lambda$ and $\nu$  all vanish. When the KN black hole is perturbed gravitationally and electromagnetically, by adopting the phantom gauge, i.e. the Maxwell scalars $\phi_0=\phi_2=0$, the perturbations are described by these Weyl scalars and spin coefficients.
Introducing
\be
\Phi_0=\Psi_0,\quad \Phi_1=\Psi_1 \rho^\ast\sqrt{2},\quad k=\frac{\kappa}{\sqrt{2}(\rho^\ast)^2},\quad
s=\frac{\sigma\rho}{(\rho^\ast)^2},
\ee
then the first set of four perturbation equations are given by \cite{Chandrasekhar:1985kt}
\begin{eqnarray}
  \left( \mc L_2 - \frac{3 i a \sin \theta}{\rho^{\ast}} \right) \Phi_0 - \left(
  \mc D_0 + \frac{3}{\rho^{\ast}} \right) \Phi_1 & = & - 2 k \left[ 3 \left( M -
  \frac{Q^2}{\rho} \right) + Q^2 \frac{\rho^{\ast}}{\rho^2} \right],\\
  \Delta_{KN} \left( \mc D_2^\dagger - \frac{3}{\rho^{\ast}} \right) \Phi_0 + \left( \mc L_{-
  1}^\dagger + \frac{3 i a \sin \theta}{\rho^{\ast}} \right) \Phi_1 & = & 2 s \left[
  3 \left( M - \frac{Q^2}{\rho} \right) - Q^2 \frac{\rho^{\ast}}{\rho^2}
  \right],\\
  \left( \mc D_0 + \frac{3}{r} \right) s - \left( \mc L_{- 1}^\dagger + \frac{3 i a \sin
  \theta}{\rho^{\ast}} \right) k & = & \frac{\rho}{\rho^{\ast 2}} \Phi_0,\\
  \Delta_{KN} \left( \mc D_2^\dagger - \frac{3}{r} \right) k + \left( \mc L_2 - \frac{3 i a \sin
  \theta}{\rho^{\ast}} \right) s & = & 2 \frac{\rho}{\rho^{\ast 2}} \Phi_1,
\end{eqnarray}
where $\rho=r+i a \cos\theta$ and $\rho^\ast=r-i a \cos\theta$. The other four perturbation equations involving $\Psi_4$, $\Psi_3$, $\lambda$ and $\nu$   will not be presented here. In this case, the gravitational perturbations is still denoted by the Weyl scalar $\Phi_0$ and the information of the electromagnetic perturbations are encoded in the Weyl scalar $\Phi_1$ and the spin coefficients $k$ and $s$. In the above equations, various operators are defined as 
\be \label{OperatorD}
\mc D_j = \partial_r + \frac{i K}{\Delta_{KN}} + 2 j\frac{r - M}{\Delta_{KN}}, \quad \mc D_j^\dagger
   = \partial_r - \frac{i K}{\Delta_{KN}} + 2j \frac{r - M}{\Delta_{KN}},
   \ee
\be
\mc L_j=
   \partial_{\theta} + P + j \cot \theta , \quad \mc L^\dagger_{j} = \partial_{\theta} - P + j\cot \theta,
 \ee
 with
\be
 P = - a \omega \sin \theta + \frac{m}{\sin \theta}, \quad K = - (r^2 + a^2) \omega + a m.
\ee

In the eikonal limit $l \gg 1$, we find that the variables can be separated by the substitutions
\be
  \Phi_0 =  R_2 (r) S_2 (\theta),\quad \Phi_1 = R_1 (r) S_1 (\theta),
\ee
\be
  k= k (r) S_1 (\theta),\quad
  s = s (r) S_2 (\theta),
\ee
where the angular functions $S_2 (\theta)$ and $S_1 (\theta)$ are the
normalized proper solutions of the equations
\be
  \mc L_{- 1}^\dagger \mc L_2 S_2 =-\mu^2 S_2,\quad
  \mc L_2 \mc L_{- 1}^\dagger S_1 =-\mu^2 S_1,
\ee
where $ \mu^2 \sim \mathcal{O} (l^2)$ . Compare with (\ref{Teuangu}) we find $\mu^2=A_2+a^2\omega^2-2am \omega$ and conclude that $S_2$ satisfies Eq. (\ref{Kerrangu}) as well in the eikonal limit $l\gg1$.

Besides, the functions $S_2 (\theta)$ and $S_1 (\theta)$ are simply related by
\be\label{S1S2}
  \mc L_2 S_2 = \mu S_1,\quad  \mc L_{- 1}^\dagger S_1 = - \mu S_2 .
\ee
Note that when $a=0$ and without taking the eikonal limit, the angular functions, $S_1$ and $S_2$, are simply related by the above formulas. However, these relations do not hold  in the Kerr case for a general $l$. As shown in the appendix \ref{EWMSW}, the angular function $S_2$ satisfies the equation (\ref{Teuangu}), but the angular function $S_1$ satisfies the equation \cite{Chandrasekhar:1985kt}
\be
(\mc L_0^\dagger\mc L_1-2a\omega \cos\theta)S_1=-\lambda_1 S_1.
\ee
Clearly, these two equations are very different. The fact that the angular functions, $S_1$ and $S_2$, are simply related by (\ref{S1S2}) is a consequence of the eikonal limit, which  is not expected to happen in general. This is the first simplification caused by taking the eikonal limit.

Taking into account of the relation between $S_1$ and $S_2$, the above perturbation equations become
\begin{eqnarray}
  \mu R_2 - \mc D_0 R_1 & = & - 2 k \left[ 3 \left( M - \frac{Q^2}{\rho} \right) +
  Q^2 \frac{\rho^{\ast}}{\rho^2} \right],\label{KNpereq1}\\
  \Delta_{KN} \mc D_2^\dagger R_2 - \mu R_1 & = & 2 s \left[ 3 \left( M - \frac{Q^2}{\rho}
  \right) - Q^2 \frac{\rho^{\ast}}{\rho^2} \right],\label{KNpereq2}\\
  \mc D_0 s + \mu k & = & \frac{\rho}{\rho^{\ast 2}} R_2,\label{KNpereq3}\\
  \Delta_{KN} \mc D_2^\dagger k + \mu s & = & 2 \frac{\rho}{\rho^{\ast 2}} R_1 .\label{KNpereq4}
\end{eqnarray}
Note that here we have taken into account that  $\mu\sim \mc O(l)$ and $K\sim \mc O(l)$, so we can safely discard  the annoying terms involving $\rho^{\ast}$ in the eikonal limit.
In fact, on the right-hand sides of the above equations, the angular dependence is still present through $\rho$ and $\rho^\ast$, which  hinders the further separation in $r$ and $\theta$. This trouble can be overcome easily in the eikonal limit. Just as the reduction made in the previous step, since $\mu\sim \mc O(l)$ and $K\sim \mc O(l)$, in (\ref{KNpereq2}) and (\ref{KNpereq4}), the coefficients of $s$ and $R_1$ on the right-hand sides  are of $\mc O(1)$, while their counterparts on the left-hand sides  are of $\mc O(l)$, so one can discard the terms on the right-hand sides and then obtains
\be
 \Delta_{KN} \mc D_2^\dagger (R_2 + k) = \mu (R_1 - s),
 \ee
 \be
 \Delta_{KN} \mc D_2^\dagger (R_2 - k) = \mu (R_1 + s).
 \ee
The fact that neither of the two equations has angular dependence implies that the complete separation in all variables has been achieved. Similarly, from (\ref{KNpereq1}) and (\ref{KNpereq3}) one has
\be
 \mc D_0 (R_1 + s) = \mu (R_2 - k),
 \ee
\be
 \mc D_0 (R_1 - s) = \mu (R_2 + k).
 \ee
Combining the above four equations, then we find a single ordinary differential equation for the gravitational perturbations
\be
 (\mc D_0 \Delta_{KN} \mc D_2^\dagger-\mu^2)R_2=0,
\ee
which explicitly gives
\be\label{KNra}
\Delta_{KN}^{-2} \frac{d}{d r} \left( \Delta_{KN}^{3}
   \frac{d^{} R_{2}}{d r^{}} \right)+ V (r) R_{2} = 0,
\ee
where
\be\label{Vrgp}
 V (r) = \frac{K^2}{\Delta_{KN}}-A_2 + 2 a m \omega - a^2 \omega^2.
 \ee
Comparing the above equation with (\ref{Kerrradial}), one can see that the two equations share the same form and the difference is only enbodied in the radial function $\Delta_{KN}$. Moreover, one can easily check that in the eikonal limit the DF equation behaves exactly the same as (\ref{Kerrangu}) and (\ref{KNra}) after the separation in $r$ and $\theta$. 

In short, for KN black holes it is feasible to separate $r$ and $\theta$ variables in the eikonal limit, due to the following two facts: one being that the two angular function $S_1$ and $S_2$ are simply related by (\ref{S1S2}), the other being that the terms involving $\rho$ and $\rho^{\ast}$ can be discarded in the eikonal limit. For RN black holes and a general $l$ , the former fact exists as well but the latter one does not. However, the spherical symmetry of the background spacetime assures the separability of the variables.  For Kerr black holes and a general $l$, even though the two angular functions are not simply related,  the term  involving $\rho$ and $\rho^{\ast}$ does not appear, which makes the the separability of the variables possible.

Different from the gravitational and electromagnetic perturbations, the scalar field is completely separable in the KN spacetime. Taking the eikonal limit, the separation of the massless scalar field in $r$ and $\theta$ as  (\ref{Scalarsep}) leads to
\be\label{KNang}
 \frac{1}{\sin \theta} \frac{d}{d \theta} \left( \sin \theta
   \frac{d S_{0}}{d \theta} \right) + \left( a^2 \omega^2 \cos^2
   \theta  - \frac{m^2}{\sin^2 \theta} + A_2 \right) S_{0} = 0,
\ee
and
\be\label{KNras}
\frac{d}{d r} \left( \Delta^{}_{KN} \frac{d R_0}{d r^{}} \right) + V (r) R_{0} = 0,
\ee
where $V(r)$ is exactly the same as that in (\ref{Vrgp}).

Obviously, the above equations differ from their counterparts in the Kerr spacetime only through the function $\Delta_{KN}$. Thus, the angular function $S_2(\theta)$ in the gravitational perturbation equation satisfies the same equation as the angular function $S_0(\theta)$ in the Klein-Gordon equation. Furthermore, similar to previous discussion around Eq. (\ref{trans}), we can easily show that Eqs. (\ref{KNra}) and (\ref{KNras}) are of the same form, i.e. the one-dimensional Schr\"odinger-like equation. From the experience of the Kerr black hole, we can obtain the QNMs from the perturbation equations (\ref{KNras}) and (\ref{KNang}) by using the WKB approximation \cite{Iyer:1986np}. Furthermore,  we expect that the QNM/geodesic correspondence (\ref{QNMGeo}) applies to the KN black hole, as we will show below.

\section{EQNM/UFPO correspondence for KN }\label{QNMG}
In this section we present the explicit relation between the high-frequencies of the QNMs and the characteristic quantities of the unstable fundamental photon orbits in the KN spacetime. Our following derivation is very close to the one in \cite{Yang:2012he}, and similarly we show that in the KN spacetime the EQNM's real frequencies are also a linear combination of the precessional and (polar) orbital frequencies, and the imaginary part of the frequencies corresponds to the Lyapunov exponent of UFPOs. 

As we know,  the frequencies of EQNMs can be calculated using the WKB approximation, with appropriate  boundary conditions. In order to obtain the  frequencies of EQNMs, one has to ensure the validity of WKB method and take into account of the boundary conditions to solve the equation of EQNMs. Next, we are not going to review the process of solving EQNMs using the WKB method  in detail \footnote{We suggest readers to see \cite{Iyer:1986np} or find a more approachable calculations in \cite{Yang:2012he} if interested in the calculation details.}, instead, we only present the necessary steps.
\subsection{QNMs from the WKB method}
Firstly, as usual, we set the complex frequency of the QNMs appearing in (\ref{separationintphi}) as
\be
\omega=\omega_R-i\omega_I,
\ee

Let us begin with the one-dimensional Schr\"odinger-like wave equation (\ref{Kerrradialst}) and its radial potential (\ref{tildeV}) with $\Delta$ replaced by $\Delta_{KN}$. The exact expression takes in this form,
\bea
\tilde{V}=\frac{\left[(r^2+a^2)\omega-am\right]^2-\Delta_{KN}\left(A_2+a^2\omega^2-2a\omega\right)}{(r^2+a^2)^2}\,.
\eea
Recall the tortoise coordinate introduced in Eq. (\ref{torr}), we find $x\to+\infty$ at $r\to\infty$ and $x\to-\infty$ at $r\to r_h$, where $r_h$ is the event horizon radius of the KN black hole. It is not hard to find that the potential $\tilde{V}$ is constant at $x=\pm\infty$, and it reaches a minimum at $x=x_0$, that is, $\partial_x \tilde{V}(x_0)=0$. For the purposes of this article, we only focus on the leading and next-to-leading orders in the WKB approximation to $\omega_R$, which is similar to \cite{Yang:2012he}, but different from \cite{Iyer:1986np}, in which the higher orders of the QNM frequency had been discussed. Thus we expand the potential $\tilde{V}$ around $r_0$ to the second order, that is,
\bea
\tilde{V}(x)=\tilde{V}(x_0)+\frac{1}{2}\partial_x^2 \tilde{V}(x_0)(x-x_0)^2\,.
\eea
In addition, as described in \cite{Yang:2012he} Sec.\Rmnum{3}.C.1 and \cite{Iyer:1986np} Sec.\Rmnum{3}.A, by analyzing the behavior of the Schrodinger-like wave function, we can know that in order to satisfy the boundary conditions of QNMs, the potential need satisfy
\bea\label{lr}
\tilde{V}(x_0)\simeq0\,.
\eea
This condition enables us to match the two WKB solutions across both of the turning points simultaneously. Thus, we finally obtain the conditions
\bea
\tilde{V}(x_0)\simeq0=\partial_x \tilde{V}(x_0)\,,
\eea
which corresponds to
\bea\label{conr}
\tilde{V}(r_0)\simeq0=\partial_r \tilde{V}(r_0)\,,
\eea
where $r_0$ and $x_0$ satisfy the Eq. (\ref{torr}). Thus, at the leading and next-to-leading orders of $\omega_R$, $\omega_R$ must satisfy
\bea\label{fcon}
\tilde{V}(r_0, \omega_R)=\partial_r \tilde{V}(r_0, \omega_R)=0\,,
\eea
from the condition (\ref{conr}).
On the other hand, a constraint has been found in \cite{Iyer:1986np} (see Eq. 1.4) to help to look for the values of the QNMs, which reads
\bea\label{iyer}
i\frac{\tilde{V}(x_0)}{\sqrt{2\partial_x^2\tilde{V}(x_0)}}=n+\frac{1}{2},
\eea
where $n$ is often referred to as the overtone number. Note that we have dropped the terms of higher orders appeared in \cite{Iyer:1986np} in our case. In fact, to derive Eq. (\ref{iyer}), Eq. (\ref{lr}) has been employed, that is, the information of Eq. (\ref{lr}) has been contained in Eq. (\ref{iyer}). To see this, we expand $\tilde{V}(x_0,\omega)$ around the point $(x_0, \omega_R)$ to the first order
\bea
\tilde{V}(x_0, \omega)=\tilde{V}(x_0, \omega_R)+\partial_\omega \tilde{V}(x_0, \omega_R)(\omega-\omega_R)\,.
\eea
and take $\omega=\omega_R-i\omega_I$, then we get
\bea\label{omegas}
\tilde{V}(x_0, \omega)=\tilde{V}(x_0, \omega_R)-i\partial_\omega \tilde{V}(x_0, \omega_R)\omega_I
\eea
We can see that due to the facts $\tilde{V}\sim\omega_R^2\sim\mc O(l^2)$ and $\partial_\omega \tilde{V}\sim\mc O(l)$ on the left hand of Eq. (\ref{iyer}), while on the right hand $n+1/2\sim \mc O(1)$. In order to save the balance of orders, we can see that the condition (\ref{lr}) is satisfied. Moreover, also as a direct consequence, we can conclude $\omega_I\sim\mc O(1)$. Using Eq. (\ref{lr}), Eq. (\ref{omegas}) can be simplified as
\bea
\tilde{V}(x_0, \omega)=-i\partial_\omega \tilde{V}(x_0, \omega_R)\omega_I\,.
\eea

Then combining with the Eq. (\ref{iyer}), we can find the final expression of $\omega_I$
\be\label{omim}
\omega_I=\left(n+\frac{1}{2}\right)\frac{\sqrt{2\partial_x^2\tilde{V}(r_0,\omega_R)}}{\partial_{\omega}\tilde{V}(r_0, \omega_R)}\,.
\ee
at the leading order of $\omega_I$. However, the calculations of $\omega$ with the WKB approximation does not end here. One can see that the potential $\tilde{V}=\tilde{V}(r, \omega, A_2)$ has three independent variables. Even though we can split $A_2$ into real and imaginary parts, that is,
\bea
A_2=A_2^R-iA_2^I\,.
\eea
and approximately take $\omega=\omega_R$ and $A_2=A_2^R$, Eqs. (\ref{fcon}) are still not enough to completely determine the value of $\omega_R$.

To close the calculations, the prerequisite is to find the relation between the separation constant $A_2$ and the position of  the turning points $r_0$ and $\omega_R$, which can be  derived from the angular equation (\ref{KNang}). Similarly, considering the boundary conditions along the angular direction and the validity of WKB method, the turning points $\theta_\pm$ divide $\theta\in[0, \pi]$ into three intervals, that is, $[0, \theta_-]$, $(\theta_-, \theta_+)$ and $[\theta_+, \pi]$. In each region, the angular function $S(\theta)$ takes different forms using the leading and next-to-leading WKB approximation. One can obtain the solutions in each region and match them at the turning points $\theta_\pm$. Then, the matching condition leads to the Bohr-Sommerfeld quantization condition \cite {Yang:2012he}, that is,
\be\label{BScond}
\int_{\theta_-}^{\theta_+}\sqrt{a^2\omega_R^2 \cos^2
   \theta  - \frac{m^2}{\sin^2 \theta} + A_2^R}\,d\theta=\left(L-|m|\right)\pi,
\ee
where $L=l+\frac{1}{2}$ and $\theta_\pm$ are the turning points of the potential of the angular equation (\ref{Kerrangu}).

\subsection{Null geodesics in the KN spacetime}
Then, we are ready to translate the above conditions into GPOs. In the KN spacetime the Hamiltonian of null particles can be separated due to the symmetries of the spacetime. One can start with the Hamilton-Jacobi equation in KN spacetime
 \be
 g^{\mu\nu}\partial_\mu S \partial_\nu S=0,
 \ee
where $S(x)$ is the principal function and $p_\mu\equiv\partial_\mu S$ is the conjugate momentum. Then the principal function can be written as
\bea\label{phs}
S(t, r, \theta, \phi)=-\E t+S_r(r)+S_\theta(\theta)+\LL \phi,
\eea
where we have used the conserved energy $\E=-p_t$ and the angular momentum $\LL=-p_\phi$ along the null geodesics.

From the separation of the Hamiltonian, one can identify another conserved quantity, viz., the  Carter constant $\Q$ \cite{Carter:1968rr}, then with the help of the conserved quantities $(\E, \LL, \Q)$ along the motion, the geodesic equation (\ref{geoeq}) in the KN spacetime can be written in the first-order form
\begin{eqnarray}
 \Sigma \dot{t} & = & \frac{r^2+a^2}{\Delta_{KN} } \left[\mc E (r^2+a^2)-\mc L a\right]-a(a \mc E \sin^2\theta-\mc L)\equiv \mc T(r,\theta),\label{timedire} \\
  \Sigma \dot{\phi} & = & -a\mc E+\frac{\mc L}{\sin^2\theta} + \frac{a [\mc E(r^2+ a^2) \E - a \LL]}{\Delta_{KN}}\equiv \mc F(r,\theta),\label{phidire} \\
\Sigma^2  \dot{\theta}^2 & = & \Q - \cos^2\theta\left(\frac{\LL^2}{\sin^2 \theta}-a^2\mc E^2\right)\equiv\Theta(\theta) ,\label{thetadot}\\
  \Sigma^2  \dot{r}^2 & = & ((r^2 + a^2) \E - a \LL)^2 -
  \Delta_{KN}  (\Q + (\LL - a \E)^2) \nn\\
  &\equiv& \R (r)=\tilde{V}(r^2+a^2)^2=\Delta_{KN} V(r),\label{rdialpoten}
\end{eqnarray}
with
\be
\Sigma=r^2+a^2\cos^2\theta,
\ee
where the dot denotes the derivative with respect to an affine parameter $\zeta$ along the null geodesics.
One can see from above equations that the difference from the case in the Kerr spacetime is completely reflected in the function $\Delta_{KN}$.
Sequentially, we can have the exact expressions of $S_r$ and $S_\theta$
\bea
S_r(r)=\bbint \frac{\pm\sqrt{\R(r)}}{\Delta_{KN}(r)}dr,\quad S_\theta(\theta)=\bbint \pm\sqrt{\Theta} d\theta,
\eea
where the ``$\bbint$'' denotes an integral along the null geodesics.
Then, from
\bea
\frac{\partial S}{\partial \E}=\frac{\partial S}{\partial\LL}=\frac{\partial S}{\partial\Q}=0,
\eea
we can obtain the relation
\bea
0&=&\partial_\Q S_r+\partial_\Q S_\theta\label{rteq},\\
t&=&\partial_\E S_r+2\partial_\E S_\theta.\label{teq}
\eea
\subsection{Correspondence with QNMs}
If we identify the principal function $S(x)$ with the phase of the geodesic equation (\ref{scalargeo}) and take into account (\ref{separationintphi}), (\ref{Kerrangu}) and (\ref{Kerrradialst}), we can immediately identify that
 \be\label{Identification}
\E=\omega_R,\quad \mc L=m,\quad \mc Q=A_2^R-m^2.
\ee
 Obviously, due to the relation of $\R(r)$ and $\tilde{V}$, (\ref{rdialpoten}), one can easily find Eq. (\ref{fcon}) is equivalent to
\be\label{fpo}
\R(r)=\R^\prime(r)=0,
\ee
where the prime denotes the derivative with respect to $r$. As we know, the photon orbits in the KN spacetime that satisfy these two equations are the FPOs, or SPOs and they are all unstable in the radial direction. Thus, we define the radius of an UFPO as $r_\fp$, and have
\bea
r_\fp=r_0\,,
\eea
where, $r_0$ is introduced in Eq. (\ref{conr}).

On the other hand, consider a nearby photon on the FPOs initially at a radius $r_{\fp}$, after $(2n+1)$ half-orbits, it advances to the larger radius $r=r_\fp+\delta r$ such that
\bea
\delta r=e^{\gamma \delta t}=e^{(n+1/2)\gamma_L \delta t}\simeq e^{(n-1/2)\gamma_L\delta t}\delta r_1,
\eea
where we introduce $\delta r_1=e^{\gamma_L\delta t}$, and $\delta t$ is the time interval for one complete orbit, that is, two halves. For one complete orbit, from Eq. (\ref{teq}), we find
\bea
\delta t=\partial_\E \delta S_r+\partial_\E \delta S_\theta,
\eea
where
\bea\label{srt}
\delta S_r=\bbint_{r_\fp}^{r_\fp+\delta r_1}\frac{\sqrt{\R(r)}}{\Delta_{KN}(r)}dr,\quad \delta S_\theta=2\bbint_{\theta_-}^{\theta_+}\sqrt{\Theta} d\theta,
\eea
with $\theta_\pm$ being the roots of $\Theta(\theta)=0$. Thus we have
\bea
\delta t=\frac{\partial_\E \R\vert_{r_{\fp}}}{\sqrt{2\R^{\prime\prime}(r_\fp)}\Delta_{KN}(r_\fp)}\log \delta r_1+\partial_\E \delta S_\theta,
\eea
where we have used
\be
\R(r)\simeq \frac{(r-r_\fp)^2}{2}\R^{\prime\prime}(r_\fp)
\ee
and from Eq. (\ref{rteq}), we find
\bea
\frac{\partial_\Q \R\vert_{r_{\fp}}}{\sqrt{2\R^{\prime\prime}(r_\fp)}\Delta_{KN}(r_\fp)}\log \delta r_1+\partial_\Q S_\theta=0.
\eea
Note that, from the matching condition in the angular direction for EQNMs, that is, Eq. (\ref{BScond}), correspondingly we have
\be
\delta S_\theta=2(L-|m|)\pi,
\ee
thus, we conclude that
\be
\partial_\E\delta S_\theta+\partial_\Q\delta S_\theta\left(\frac{d\Q}{d\E}\right)=0,
\ee
and therefore, we would have
\be
\frac{1}{\sqrt{2\R^{\prime\prime}(r_\fp)}\Delta_{KN}(r_\fp)}\left[\partial_\E \R\vert_{r_{\fp}}+\partial_\Q \R\left(\frac{d\Q}{d\E}\right)\bigg|_{r_\fp}\right]\log \delta r_1=\delta t,
\ee
thus we have
\be
\gamma_L=\frac{\log\delta r_1}{\delta t}=\frac{\sqrt{2\R^{\prime\prime}(r_\fp)}\Delta_{KN}(r_\fp)}{\partial_\E \R\vert_{r_{\fp}}+\partial_\Q \R\left(\frac{d\Q}{d\E}\right)\big|_{r_\fp}},
\ee
thus we finally find
\be
\gamma=\left(n+\frac{1}{2}\right)\frac{\sqrt{2\R^{\prime\prime}(r_\fp)}\Delta_{KN}(r_\fp)}{\partial_\E \R\vert_{r_\fp}+\partial_\Q \R\left(\frac{d\Q}{d\E}\right)\big|_{r_\fp}},
\ee
Next, from Eqs. (\ref{fcon}) and (\ref{rdialpoten}), we find
\bea
\R=\R^\prime=\tilde{V}=\tilde{V}^\prime=0,\quad\quad\text{at}\quad\quad r=r_\fp=r_0\,,
\eea
then, combining with the definition of the tortoise coordinate \ref{torr}, we have
\bea
\partial_x^2\tilde{V}=\frac{\Delta_{KN}^2}{(r^2+a^2)^2}\partial_r^2\tilde{V}=\frac{\Delta_{KN}^2}{(r^2+a^2)^4}\partial_r^2[\tilde{V}(r^2+a^2)^2]=\frac{\Delta_{KN}^2}{(r^2+a^2)^4}\R^{\prime\prime}\,,
\eea
at $r=r_\fp=r_0$. On the other hand, the radial potential can be seen as $\R=\R(r, \E)$, thus we have
\bea
\partial_\E \R\vert_{r_\fp}+\partial_\Q \R\left(\frac{d\Q}{d\E}\right)=\partial_\E\R\,,
\eea
and then we get
\bea
\partial_\omega\tilde{V}=\frac{\partial_\E\R}{(r^2+a^2)^2}\,.
\eea
Thus, we obtain
\bea
\gamma=\left(n+\frac{1}{2}\right)\frac{\sqrt{2\R^{\prime\prime}(r_\fp)}\Delta(r_\fp)}{\partial_\E \R\vert_{r_\fp}+\partial_\Q \R\left(\frac{d\Q}{d\E}\right)\big|_{r_\fp}}=\left(n+\frac{1}{2}\right)\frac{\sqrt{2\partial_x^2\tilde{V}(r_0,\omega_R)}}{\partial_{\omega}\tilde{V}(r_0, \omega_R)}\,,
\eea
compared with Eq. (\ref{omim}), we have
\be\label{omei}
\omega_I=\gamma=\left(n+\frac{1}{2}\right)\gamma_L,
\ee
where $\gamma_L$ is known as the  Lyapunov exponent. Up to now, we can see that with the two additional matching conditions of EQNMs, one can confirm that the corresponding GPOs are UFPOs.

Moreover, as in \cite{Ferrari:1984zz,Yang:2012he}, we can also introduce two frequencies associated with individual spherical photon orbits, viz., the orbital and precessional frequencies, and connect them with the real part of the QNM frequency. The $\theta$-frequency is defined as
\bea
\orr\equiv2\pi/\delta t\,,
\eea
 with $\delta t$ being the the time interval of a complete $\theta$-cycle, and
\bea
\pr\equiv \delta\phi_{\rm{prec}}/\delta t\,,
\eea
is the $\phi$-frequency, where $\delta \phi_{\rm{prec}}$ is the difference between the angle the particle accumulate in the azimuthal direction during a complete $\theta$-cycle and $\pm2\pi$, i.e. $\delta\phi_{\rm{prec}}= \delta\phi-2\pi \cdot {\rm sgn}(m)$, here we use $m>0$ to denote a corotating orbit while $m<0$ denotes a  counterrotating orbit. In addition,  $\delta t$ and $\delta \phi$ can be computed out along the null geodesics. From Eqs. (\ref{timedire}), (\ref{phidire}) and (\ref{thetadot}), we have
\bea
\frac{dt}{d\theta}=\frac{\mc T(r,\theta) }{\sqrt{\Theta}},\quad \frac{d\phi}{d\theta}=\frac{\mc F(r,\theta)}{\sqrt{\Theta}}\,,
\eea
And then we can obtain
\be
\delta t=2\int_{\theta_-}^{\theta^+}\mc T(r,\theta) \frac{d\theta}{\sqrt{\Theta}},
\ee
and
\be
\delta\phi=2\int_{\theta_-}^{\theta^+}\mc F(r,\theta) \frac{d\theta}{\sqrt{\Theta}},
\ee

Recall that $S(t, r, \theta, \phi)$ is the phase of the photons, $S$ should be unchanged during the time interval of a complete orbit $\delta t$ near $r=r_\fp$. Thus, from Eq. (\ref{phs}) we have
\bea\label{deltaph}
-\E\delta t+\delta S_r+\delta S_\theta+\LL\delta\phi=0\,,
\eea
where $\delta S_r$ and $\delta S_\theta$ have been introduced in Eq. (\ref{srt}). Obviously, we can drop the term $\delta S_r$, since $\R(r_\fp)=\R^{\prime\prime}(r_\fp)=0$.

Now Eq. (\ref{deltaph}) can be rewritten as
\bea
-\omega_R \delta t+\LL \delta\phi+\delta S_\theta=0
\eea
then we can find
\bea
\omega_R=\frac{m\delta\phi+\delta S_\theta}{\delta t}&=&\frac{m\left[\delta\phi_{\rm{prec}}+2\pi{\rm sgn}(m)\right]+2\pi(L-|m|)}{\delta t}\nn\\
&=&\frac{m\delta\phi_{\rm{prec}}+2\pi L}{\delta t}=L\left(\orr+\frac{m}{L}\pr\right)\,.
\eea

Combining with Eq. (\ref{omei}), we show
\be
\omega=L\left(\orr+\frac{m}{L}\pr\right)-i\left(n+\frac{1}{2}\right)\gamma_L,
\ee
is valid for the QNMs of the KN black hole as well. Note that this formula has the same form as Eq. (1.3) in \cite{Yang:2012he}, which was derived for the QNMs of Kerr black holes.

\section{Conclusions and discussion}\label{CD}

In this paper, we studied the EQNM/UFPO correspondence \cite{Ferrari:1984zz,Cardoso:2008bp,Yang:2012he} for the  black holes in the Einstein-Maxwell theory.  The explicit content of the EQNM/UFPO correspondence for the Kerr black holes was well explored in \cite{Yang:2012he}. We tried to shed new light on this correspondence. We found that in the eikonal limit both the Teukolsky equation and the (separated) massless Klein-Gordon equation in the Kerr spacetime can be turned into the same one-dimensional Schr\"odinger-like wave equation. This simple fact plays an essential role in setting up the EQNM/UFPO correspondence. Since the massless Klein-Gordon equation in the eikonal limit can also be interpreted as the null geodesic equation, the fact implies that the EQNMs must correspond to some particular GPOs. Employing the WKB method, it turns out that the boundary conditions  in the radial and angular direction on the EQNMs is equivalent to the requirements that GPOs must be UFPOs, or homoclinic null geodesics\cite{Yang:2012he}. Consequently, the imaginary part of the  frequency of the QNM of the overtone number $n$ is related to the Lyapunov exponent of the photon trajectory circling $(2n+1)$ half-orbits

Moreover, we studied the EQNM/UFPO correspondence for the Kerr-Newman black hole. We showed  that in the eikonal limit the gravitational and electromagnetic perturbations of the
Kerr-Newman black hole are naturally decoupled, from which a single one-dimensional
Schr\"odinger-like equation encoding the QNM spectrum can be derived. We then showed that the analog of the Teukolsky equation and the (separated) massless Klein-Gordon equation in the Kerr-Newman spacetime are of the same form when taking the
eikonal limit. This allows us to set up the correspondence between the EQNM and UFPOs.  In particular, similar to the Kerr case (\ref{QNMGeo}) , the quasinormal mode's real frequency is a linear combination of the precessional and (polar) orbital frequencies, and the imaginary part of the frequency is proportional to the Lyapunov exponent of the spherical photon orbit.

In the literatures there have been found some examples that EQNMs and UFPOs do not match for the black holes in AdS spacetime \cite{Cardoso:2008bp} and the black holes in modified theories of gravity \cite{Konoplya:2017wot}. Our study in this paper may give some insights on these problems. For the former, one can see that the equation of EQNM still shares the same as the one of null geodesic, however, the boundary condition along the radial condition for EQNMs has changed, so that the EQNM/UFPO correspondence is broken. Nevertheless, it is possible that EQNM could correspond to some other GPO than UFPO. It would be interesting to study this possibility further. For the latter, even though the boundary conditions of EQNM remain unchanged,  due to the presence of higher order derivative terms, the equation of EQNM is different from the null geodesic equation.

\section*{Acknowledgments}
The work is in part supported by NSFC Grant  No. 11735001. MG is also funded by China Postdoctoral Science Foundation Grant No. 2020T130020. PCL is also funded by China Postdoctoral Science Foundation Grant No. 2020M670010.
\appendix
\section{A brief review of the EQNM/UFPO correspondence}\label{EWMSW}

In this section, we would like to briefly review of the EQNM/UFPO correspondence for the Kerr spacetime.  We show that the equations of EQNMs are the same as the ones of free moving photons, which is a necessary condition for EQNM/UFPO corresponce. As the first step, let us introduce the geometric optics approximation, sometimes also called the eikonal limit, to electromagnetic waves (EWs) \cite{Maggiore:1900zz} and massless scalar waves (MSWs) \cite{Misner:1974qy, Yang:2012he}, respectively. They are actually equivalent under the geometric optics approximation.

Let us begin with the gauge field $A_\mu$ which satisfies the source-free Maxwell equations
\bea\label{smax}
\nabla_\mu F^{\mu\nu}=0,
\eea
where $F_{\mu\nu}=\partial_\mu A_\nu-\partial_\nu A_\mu$.
Imposing the Lorenz gauge $\nabla_\mu A^\mu=0$, Eq. (\ref{smax}) can be rewritten as
\be \label{Maxwelleq}
\nabla^{\rho} \nabla_{\rho} A^{\mu} - R^{\mu}_{\,\,\rho} A^{\rho} = 0,
\ee
where we have utilized the Ricci identity, i.e. $\nabla_\rho\nabla_\nu A^\rho-\nabla_\nu\nabla_\rho A^\rho=R_{\nu\rho}A^\rho$ and $R_{\mu\nu}$ is the Ricci tensor. The validness of the geometric optics approximation requires the wavelength $\lambda$ is much smaller than the other length scales in the problem, which can be uniformly denoted by $L$, such as the curvature radius of the background metric and the typical length scale of variation of the amplitude, polarization or the wavelength of the electromagnetic field.

Under the geometric optics approximation $\lambda\ll L$, we can write
\be
A^\mu(x)=a^\mu(x) e^{i S(x)},
\ee
where the phase $S(x)$ changes on the scale $\lambda$ and is rapidly varying, while the amplitude changes only on the scale $L$ and is slowly varying.  Since $ R^{\mu}_{\,\,\rho} A^{\rho}=\mc O(L^{-2})$ while $\nabla^{\rho} \nabla_{\rho} A^{\mu}=\mc O(\lambda^{-2})$, then
up to the leading and the next-to-leading order in $\lambda/L$ we can neglect
$ R^{\mu}_{\,\,\rho} A^{\rho}$, and the Maxwell equation Eq.(\ref{Maxwelleq}) is simply
\be \label{Meq2}
\nabla^{\rho} \nabla_{\rho} A^{\mu} = 0.
\ee
Defining the wavevector $k_\mu\equiv \partial_\mu S$, then from the Lorenz gauge we obtain $k_\mu a^\mu=0$.
From Eq.(\ref{Meq2}), to the lowest order, we get
\be\label{LOeq}
g_{\mu\nu} k^\mu k^\nu=0,
\ee
which is known as the {\em eikonal equation}, and one can show that it is equivalent to the
geodesic equation
\be\label{geoeq}
k^\mu \nabla_\mu k_\nu=0.
\ee
From the point of view of the Hamilton-Jacobi formalism, the phase $S(x)$ could be interpreted as the principal function, and the eikonal equation just corresponds to the Hamilton-Jacobi equation for massless particles.

To the next-to-leading order in $\lambda/L$, the Maxwell equation (\ref{Meq2}) gives
\be
2k_\rho \nabla^\rho a^\mu+(\nabla^\rho k_\rho)a^\mu=0,
\ee
 which, in terms of the scalar amplitude $a\equiv(a^\mu a_\mu)^{1/2}$, can be written as
 \be\label{amplitude}
 2k^\mu \partial_\mu \log a+\nabla_\mu k^\mu=0.
 \ee
The fundamental equations (\ref{LOeq}) and (\ref{amplitude}) contain the necessary information about the propagation of a null geodesic in curved spacetime.

In fact, the fundamental equations can also be derived from the Klein-Gordon equation for a massless scalar field \cite{Yang:2012he},
\be\label{KGeq}
\nabla^2\Phi(x)=0.
\ee
Similarly, after writing
\be\label{scalargeo}
\Phi(x)=u(x) e^{i S(x)},
\ee
and setting $k_\mu\equiv\partial_\mu S$, by requiring the phase $S(x)$ changes on the scale $\lambda$, while the amplitude $u(x)$ changes
only on the scale $L$, from Eq. (\ref{KGeq}) we can obtain the following equations
\be\label{scalargeom}
g_{\mu\nu}k^\mu k^\nu=0,\quad 2k^\mu\partial_\mu \log u+\nabla_\mu k^\mu=0
\ee
at the leading order and the next-to-leading order in $\lambda/L$, respectively. Comparing them with  Eqs. (\ref{LOeq}) and (\ref{amplitude}), we find they have the same forms by identifying $u(x)=a(x)=(a^\mu a^\ast_\mu)^{1/2}$. This means that  it does not matter which kind of field will be used in the geometric optics approximation, as all of them should be described by null geodesics. Therefore, we will use MSW equations in the following discussion.

It is known that for the Schwarzschild, the Reissner-Nordstrom and the Kerr black holes, the small  perturbations can be described by a set of  linear second-order partial differential equations, which can be separated completely \cite{Chandrasekhar:1985kt}. Formally, the perturbations of the stationary spacetime can be denoted by a field expressed as
\be\label{separationintphi}
 \Psi =\sum_{l,m}\int d\omega e^{-i \omega t} e^{i m\phi} S_{\omega lm}(\theta)R_{\omega lm} (r),
 \ee
 where $\omega$ is the frequency, $l$ and $m$ are the angular multipoles, due to the translational and rotational symmetry of the spacetime.

According to the behavior  under the parity operations, the gravitational perturbations of the
Schwarzschild black hole can be classified and decoupled into the axial and the polar sectors. The study of the axial sector was initiated by Regge and Wheeler \cite{Regge:1957td}, and the polar sector was analyzed by Zerilli \cite{Zerilli:1971wd}. In \cite{Chandrasekhar:1985kt} Chandrasekhar had shown that these two sectors can be transformed into each other and yield identical spectrum of qusinormal modes, i.e. the two sectors are {\em isospectral}.

The approach taken by Regge, Wheeler and Zerilli is to study directly the perturbations of the metric via the linearized Einstein's equation  about the background spactime. However, one can also study the perturbations in  the Newman-Penrose (NP) formalism \cite{Newman:1961qr}. The latter avenue is particularly suitable for the study of the gravitational perturbations of the Kerr black hole. Via the NP formalism, Teukolsky derived the equations describing the perturbations of Kerr black hole, which are completely separable into ordinary differential equations (called the Teukolsky equations) \cite{Teukolsky:1973ha}. Taking $a\to 0$ limit, the Teukolsky equations naturally reproduce the equations of the gravitational perturbations of the Schwarzschild black hole.

For the Kerr black holes, the gravitational perturbations are encoded by the linearized Weyl scalars $\Psi_0$ and $\Psi_4$, which are gauge invariant under infinitesimal diffeomorphisms. Here for simplicity  we only focus on $\Psi_0$ and similar discussion can be made for $\Psi_4$. Following \cite{Chandrasekhar:1985kt}, it can be  separated in $r$ and $\theta$,
 \be
 \Psi_0(r,\theta)=R_2(r)S_2(\theta),
 \ee
 and the perturbation  equations reduce to
 \be
 (\Delta \mc D_1 \mc D_2^\dagger+6i\omega r)R_2=\bar{\lambda}_2 R_2,
 \ee
 \be\label{Teuangu}
 (\mc L^\dagger_{-1}\mc L_2-6 a \omega\cos\theta)S_2=-\bar{\lambda}_2 S_2,
 \ee
where $\Delta=r^2-2M r+a^2$ and various operators share the same form as (\ref{OperatorD}) but with $\Delta$ replacing $\Delta_{KN}$.

Note that although the above equations derived by Chandrasekhar are equivalent to the ones by Teukolsky \cite{Teukolsky:1973ha}, the separation constant is different. The two separation constants are related by $\bar{\lambda}_2=A_2+a^2\omega^2-2am \omega$, where $A_2$ is the one used by Teukolsky. In the following we prefer to use $A_2$ instead of $\bar{\lambda}_2$, since the former one appears mostly in the literatures.

Taking the eikonal limit $l\gg1$, these two equations become
\be \label{Kerrangu}
\frac{1}{\sin \theta} \frac{d}{d \theta} \left( \sin \theta
   \frac{d^{} S_{2}}{d \theta} \right) + \left( a^2\omega^2 \cos^2
   \theta  - \frac{m^2}{\sin^2 \theta} + A_2\right) S_{2} = 0,
\ee
and
\be \label{Kerrradial}
\Delta^{-2} \frac{d}{d r} \left( \Delta^{3}
   \frac{d^{} R_{2}}{d r^{}} \right) + V (r) R_{2} = 0,
\ee
where
\be
 V (r) = \frac{K^2}{\Delta}-A_2 + 2 a m \omega - a^2 \omega^2.
 \ee
Note that the limit $l\gg1$ and the high frequency limit, i.e. the geometric optics limit $\omega\gg1$ are essentially independent of each other.  Since the frequency appearing in the above two equations is the eigenvalue to be determined, as a consequence we have $A_2\sim \mc O(l^2)$, $\omega\sim O(l)$ and $m\sim O(l)$.
For $a=0$, the solution of the angular equation (\ref{Kerrangu}) is just the Legendre function $P_l$ with $A_2=l(l+1)$.
Besides, via the transformation
\be
\tilde{R}_2=\sqrt{r^2+a^2} \Delta R_2,
\ee
and the tortoise coordinate
\be\label{torr}
dx=\frac{r^2+a^2}{\Delta} dr,
\ee
the radial equation (\ref{Kerrradial}) becomes  the one-dimensional Schr\"odinger-like wave equation
\be\label{Kerrradialst}
 \frac{d^2}{d x^2} \tilde{R}_{2} +\tilde{V}  \tilde{R}_{2} = 0,
 \ee
where
\be\label{tildeV}
 \tilde{V} \simeq \frac{\Delta}{(r^2 + a^2)^2} V.
 \ee

On the other hand, the scalar field is completely separable in the Kerr spacetime. Taking the eikonal limit, from the Klein-Gordon equation (\ref{KGeq}) the separation of the massless scalar field in $r$ and $\theta$, i.e.
\be\label{Scalarsep}
\Phi(r,\theta)=R_0(r) S_0(\theta),
\ee
leads to
\be\label{Scalarangu}
 \frac{1}{\sin \theta} \frac{d}{d \theta} \left( \sin \theta
   \frac{d S_{0}}{d \theta} \right) + \left( a^2 \omega^2 \cos^2
   \theta  - \frac{m^2}{\sin^2 \theta} + A_2 \right) S_{0} = 0,
\ee
and
\be \frac{d}{d r} \left( \Delta^{} \frac{d R_0}{d r^{}} \right) + V (r) R_{0} = 0,
\ee
where
\be V (r) = \frac{K^2}{\Delta} + 2 a m \omega - a^2
   \omega^2 -A_2.
   \ee
Clearly, the angular equation for the scalar field (\ref{Scalarangu}) has the same form as that of the gravitational perturbations (\ref{Kerrangu}), which means we have $\bar{\lambda}_2=\bar{\lambda}_0$. Moreover, by using the tortoise coordinate and via the transformation
\be\label{trans}
\tilde{R}_0=\sqrt{r^2+a^2} R_0,
\ee
the radial equation of the scalar field can also be transformed into the  standard one-dimensional Schr\"odinger-like equation
\be
 \frac{d^2}{d x^2} \tilde{R}_{0} + \tilde{V}  \tilde{R}_{0} = 0,
 \ee
where $\tilde{V}$ is exactly the same as that in (\ref{Kerrradialst}).

From the above discussions we arrive at the conclusion that in the eikonal limit the equations describing the gravitational and the scalar perturbations possess essentially the same form\footnote{In higher dimensions, the scalar, the electromagnetic and the gravitational perturbations of static black holes in Einstein's gravity have the same behavior in the eikonal limit \cite{Kodama:2003jz,Ishibashi:2003ap,Kodama:2003kk}.}. Moreover,  as pointed out in \cite{Yang:2012he} considering the infalling boundary condition at the horizon and the outgoing boundary condition at infinity and the validity of the WKB method, two additional matching conditions  should be imposed on the radial and angular directions respectively to find the eigenvalues of EQNMs. The Corresponding two boundary conditions imposed  on GPOs restrict the photon orbits to be UFPOs. This establishes  the EQNM/UFPO correspondence,  as encoded in the relation \eqref{QNMGeo}.

\providecommand{\href}[2]{#2}

\end{document}